\documentclass[epsf,twocolumn,showpacs,preprintnumbers]{revtex4}
\usepackage{graphics}
\usepackage{graphicx}
\usepackage{dcolumn} 
\usepackage{bm}
\usepackage{epsfig}
\begin{document}
\title{Charge and Spin Ordering in Insulating 
  Na$_{0.5}$CoO$_2$: \\
  Effects of Correlation and Symmetry } 
\author{K.-W. Lee and W. E. Pickett} 
\affiliation{Department of Physics, University of California, Davis, CA 95616}
\date{\today}
\begin{abstract}
Ab initio band theory including correlations due to intra-atomic
repulsion is applied to study
charge disproportionation and charge- and spin-ordering in insulating
Na$_{0.5}$CoO$_2$.  Various ordering patterns (zigzag and two striped) for
four-Co supercells are analyzed before focusing on the
observed ``out-of-phase stripe'' pattern of antiferromagnetic Co$^{4+}$
spins along charge-ordered stripes.
This pattern relieves frustration and shows distinct analogies with the
cuprate layers: a bipartite lattice of antialigned spins, with axes at
90$^{\circ}$ angles.  Substantial distinctions with cuprates are 
also discussed, including the
tiny gap of a new variant of ``charge
transfer'' type within the Co $3d$ system.  
\end{abstract}
\maketitle

The Na$_x$CoO$_2$ system, which forms the basis for a quasi-two-dimensional
transition metal oxide superconductor (T$_c$ = 4.5 K) 
when hydrated,\cite{takada}
shows a wide variety of unexplained behaviors in the 
accessible range $0<x\le 1$.
Counterintuitively (considering it is a 2D transition metal oxide) it 
shows {\it uncorrelated} behavior in the normal state for 
$x < 0.5$\cite{PRB04} 
(superconductivity arises when
$x\approx 0.3$ samples are hydrated).   Then, rather unexpectedly for a hole-doped
band insulator, it displays correlated behavior for $x > 0.5$ 
including an enhanced
linear specific heat coefficient and local moments (Curie-Weiss susceptibility).
Both of these regimes are metallic.  Precisely at $x$=0.5, however,
it evolves through Na ion ordering, charge ordering, and magnetic
ordering transitions to attain 
a ground state that is insulating\cite{foo} with a very small gap (few tens of meV).

\begin{figure}
\rotatebox{-90}{\resizebox{4.0cm}{8cm}{\includegraphics{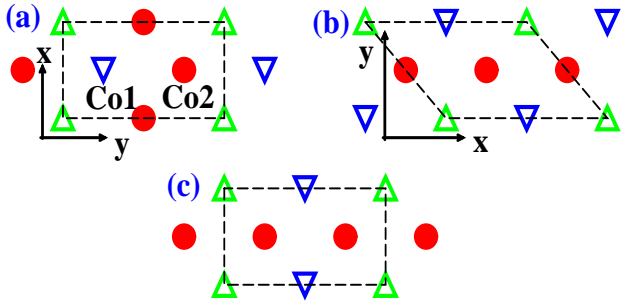}}}
\caption{(color online)Charge and spin ordering of (a)zigzag (ZZ)
 and (b) stripe (OP-ST) patterns suggested by Ga\v{s}parovi\'{c} {\it et al.}
 and Yokoi {\it et al.}, and 
(c) in-phase stripe in the Co layer.
 In the OP-ST model (b) the Co2 ions lies at a site of in-plane inversion
symmetry and is neighbored symmetrically by $\uparrow$ and $\downarrow$ spins.
 Triangle and filled circle denote magnetic (Co1) and nonmagnetic (Co2)
 cobalts, respectively, and
 the direction of the triangles indicates spin orientation.
 In the calculations, Na lies above Co2.}
\label{cell}
\end{figure}

Much of the interest in this system lies in the triangular arrangement of the
Co ions, and the expectation that the system should be addressable in terms
of nonmagnetic Co$^{3+}$ and spin-half Co$^{4+}$ ions.  Ordering phenomena,
whether charge, spin, or orbital, acquires a different character on a triangular
lattice\cite{honerkamp,ogata,bulut} 
than on the heavily studied square lattice of the cuprate superconductors.
The system becomes
magnetically ordered at $x \geq 0.75$ (antialigned stacking of ferromagnetic
layers\cite{MPI}).
The insulating state was discovered by Foo {\it et al.}\cite{foo}, 
who presented electron
diffraction data indicating robust Na ion ordering (in an orthorhombic
four-Co supercell\cite{huang,zandbergen} whose cell shape is shown in 
Fig. \ref{cell}.  The Na ordering
persisted to above room temperature, and it was suggested that Na ordering
could be coupled to charge (hole) ordering.  The zigzag Na order they inferred, 
involving equally the two distinct types of Na sites, 
was confirmed by Yang {\it et al.}\cite{yang}, and calculations by Zhang
{\it et al.} concluded that this ordering is favored
because it minimizes the Coulomb interactions between 
the Na$^+$ ions.\cite{zhang}  

As the temperature is lowered, a kink in the in-plane susceptibility
$\chi_{ab}$ at T$_{c1}$=88 K signals antiferromagnetic (AFM) ordering of some
Co spins.\cite{foo,yokoi,mit}    
Infrared reflectivity studies\cite{hwang,wang,lupi} detect a gap 
of $\sim$15 meV opening below
T$_{c2} =$ 52 K, where Foo {\it et al.}
observed\cite{foo} the onset of insulating behavior in the 
resistivity $\rho(T)$. 
T$_{c2}$ has been called the charge-ordering temperature but there is also
additional magnetic rearrangement, signaled by a kink in 
$\chi_c$.\cite{yokoi,mit}
At T$_{c3}$=27 K Ga\v{s}parovi\'{c} {\it et al.} observed an additional kink
in $\chi_{ab}$ with no signature in $\rho(T)$; this is the temperature
where Foo {\it et al.} had observed structure in $\rho(T)$ reflecting more
highly insulating behavior.\cite{foo}   Unlike the upper two transitions, there is
no entropy change\cite{huang} at T$_{c3}$.
The interpretation of
this onset of insulating behavior was suspected to be charge 
ordering,\cite{foo,hwang} inviting neutron diffraction studies.  The
two studies reported to date\cite{yokoi,mit} 
confirm that (1) there are two types of Co ions ({\it i.e.} 
charge disproportionation), one consistent
with spin half but with a reduced ordered moment (0.25-0.34 $\mu_B$), 
the other spin being much smaller,
and (2) AFM ordering below T$_{c1}$=88 K seems to be 
the type shown in Fig. \ref{cell}(b). 

First principles local density approximation (LDA) calculations by Singh
at $x$=0.5 predict\cite{singh1} ferromagnetic (FM) Co layers to be 
much more stable than a simple
AFM arrangement, and treating the Na ions explicitly does not
change this conclusion.\cite{li}  A crucial development occurred with the
discovery that, by using the correlated LDA+U method, explicit charge
disproportionation occurs\cite{PRB04} above a critical value $U_{cr}$.  
Subsequent ordering of the holes
then can relieve the frustration on the triangular lattice, whereupon it was
found that AFM order became favored over FM at $x$=0.5.\cite{PRL05}  

The questions of type(s) of order, mechanism of ordering, and character of
the insulating state have begun to be clarified by the data discussed
above.  A quantitative understanding of the behavior is likely to 
require accounting for: the multiband $t_{2g}$ system with symmetry broken
down to singlet $a_g$ + doublet $e_g^{\prime}$ by the hexagonal ligand field; 
the triangular, non-bipartite lattice that
frustrates AFM ordering and provides several nearly degenerate possibilities
for charge order\cite{zhang,mizokawa,choy}; correlation effects 
strong enough to drive charge
disproportionation  but small enough to leave a tiny charge gap.  

We have addressed these questions using the same methods\cite{fplo} as for our
previous studies on this system,\cite{PRB04,PRL05}  with attention to Brillouin
zone sampling (up to 312 k-points in the irreducible zone). Specifically, we utilize
the disproportionated states provided by the LDA+U approach to address, for
the three superlattice symmetries shown in Fig. \ref{cell}, the energetics,
the relative orientations and  magnitudes of the magnetic moments, and characteristics
of the electronic structure in the insulating phase.  These Co orderings 
correspond to (a) zigzag (ZZ), (b) out-of-phase stripe (OP-ST), and 
(c) in-phase
stripe (IP-ST).  Note that the Na ion zigzag phase is not the same as
this Co ZZ order; the Na zigzag involves one site on top of Co and another site not
on top of any Co that is ``less zigzag'' than this Co ZZ order.   
There is some controversy of the effect of Na order: Na-Na repulsion accounts for
the preferred order \cite{zhang} without further mechanisms; Na$^+$ order and 
Co$^{3+}$/Co$^{4+}$ order are coupled \cite{meng}; or the observed 
Co$^{3+}$/Co$^{4+}$ order can be obtained from a single-band extended Hubbard
model \cite{choy} without any further complications.
Li {\it et al.}\cite{li} used the observed Na superstructure \cite{huang}
and included O ion relaxation but did not compare the results
with simpler Na arrangements \cite{geck}.
We do not address this question here but note that the Na ordering we adopt,
used in previous work,\cite{PRL05}
serves to break the the symmetry of the Co sites.
Specifically, the
Na ions sit above the Co2 site (which becomes the nonmagnetic Co$^{3+}$ site).


Our attempts to obtain an AFM state for the IP-ST model of Fig. \ref{cell}(c)
converged to a FM state (or nonmagnetic, if AFM symmetry was 
enforced).\cite{explain}
Thus we will consider only the model given in Fig. \ref{cell}(b) 
as a stripe (ST) pattern in this paper.
Within LDA, FM is favored over AFM for both ZZ and ST patterns, as for 
all other values of $x$.\cite{singh1}  The energy difference is substantial
for ZZ (300 meV/Co) but surprisingly small for ST (8 meV/Co).  This favoring
of FM by LDA confirms the need for including effects of correlation, as we do by
applying the LDA+U method.  As emphasized previously, results depend on the 
value of $U$ and it is necessary to determine the appropriate value.  This
is rather straightforward for $x$=0.5: there should be disproportionation,
charge- and spin-ordering (AFM) and a very small gap.  We first review
behavior versus the repulsion strength $U$. Hund's rule J=1 eV is kept fixed.

\begin{figure}
\rotatebox{-90}{\resizebox{7cm}{7cm}{\includegraphics{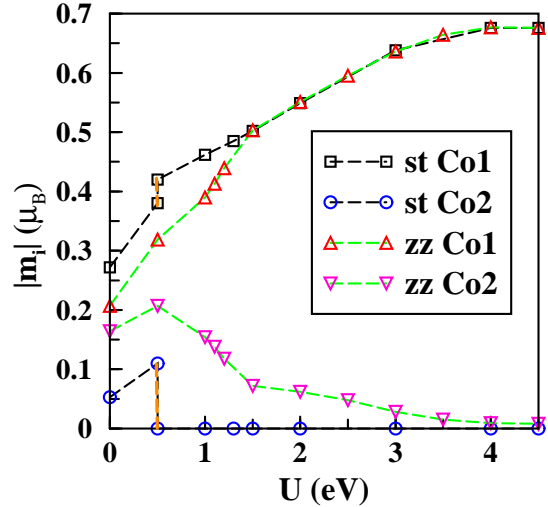}}}
\caption{(color online)Effect of $U$ on magnitude of the Co local magnetic
 moments $m_i$ in the stripe and zigzag patterns of AFM Na$_{0.5}$CoO$_2$.
  At U$_c$=1.5 eV, charge disproportionation Co1$\rightarrow$Co$^{4+}$ (S=1/2) and
  Co2$\rightarrow$Co$^{3+}$ (S=0), occurs with gap opening.}
\label{um}
\end{figure}

The effect of $U$
is evident in the calculated Co moments,
displayed in Fig. \ref{um}.
At $U$=0 ({\it i.e.} LDA level) the effects of symmetry (determined by the
Na placement) is already strong.  The Co1 and Co2 moments are nearly equal
for ZZ, while there is already almost negligible moment on Co2 for ST.
This difference reflects the higher symmetry of the Co2 ion in the ST pattern
of Fig. \ref{cell}(b): it is surrounded symmetrically (in-plane inversion)
by two Co1$\uparrow$,
two Co1$\downarrow$, and two nonmagnetic Co2.
Increasing $U$, Co1 magnetic moments in both patterns increase monotonically 
and become identical at and above $U$=1.5 eV.  
The low-spin Co2 magnetic moments show a much greater difference between 
the two patterns. For ST it becomes immediately (by $U$=0.5 eV) zero, while
for ZZ there is at $U$=0.5 eV what might be identified as the charge
disproportionation transition, but beyond this point the Co2 
moment simply decreases
monotonically, never becoming identically zero. From Fig. \ref{cell}(a) 
the lack of symmetry in ZZ is clear: although surrounded by two $\uparrow$ and
two $\downarrow$ Co1 spins, and two low-spin Co2 ions, there 
is no in-plane inversion,
so a moment is allowed.  While the Co1 and Co2 ions are clearly disproportionated
in Fig. \ref{um}, the charge difference is only $\sim$0.2 e (this 
difference is 0.02 e smaller for ST than ZZ).

\begin{figure}
\resizebox{8cm}{5cm}{\includegraphics{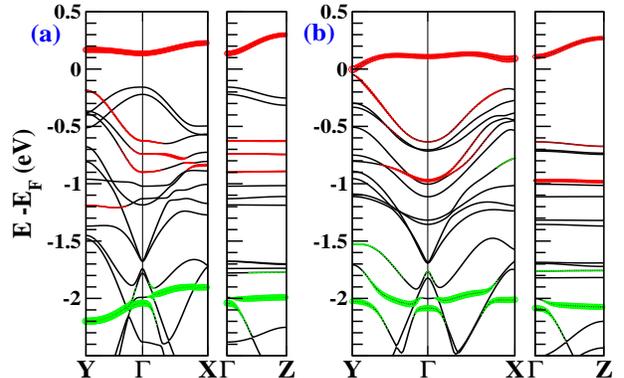}}
\caption{(color online) View of the AFM band structures in the $t_{2g}$ manifold
 at U=2 eV for (a) the zigzag (ZZ) and (b)the stripe (ST) patterns.
 The plot is along perpendicular ($\Gamma$-Y) and
 parallel ($\Gamma$-X) directions for the each chain.
 Z denotes the zone-boundary point along $<001>$ direction.
 The thickened lines emphasize the bands with strong $a_g$ character
 for each spin of a magnetic Co1.
 The energy zero lies in the gap.}
\label{band}
\end{figure}

The gap opens, for both patterns, at $U$=$U_{cr}$=1.5 eV.  It is noteworthy
that the disproportionation had occurred already at smaller U (see Fig.
\ref{um}); thus we find here a richer behavior than in our previous
studies with smaller cells or different Na concentrations, where 
disproportionation/ordering had coincided with gap opening.  Such a difference
was also obtained for a similar supercell by Li {\it et al.}, who however
found a somewhat larger value of U$_{cr}$.\cite{li}  This critical interaction
strength coincides with a Mott-like transition in the Co1 $a_g$ states, 
with upper and lower
Hubbard bands separated by 2.2 eV as shown in Fig. \ref{band}.  
The distinctive $a_g$ state is prominent already within LDA, arising from
symmetry breaking due to the hexagonal ligand field
($t_{2g}\rightarrow a_g + e_g^prime$), and the gap opening at $U_{cr}$
constitutes the band description of an orbitally selective Mott transition.
Almost independent
of $U$, the ZZ pattern is favored over ST by the very small value of 22 meV/Co;
given the fact that the Na zigzag arrangement has been simplified in
our calculations, we can
conclude that these two patterns have no significant difference in energy.

The upper Hubbard band in the insulating state is quite flat,
and in fact shows more dispersion
perpendicular to the layers (200 meV) than within the layers. 
The band structures are pictured along directions parallel to 
($\hat x$), and perpendicular
to ($\hat y$), the ZZ or ST chains of Co ions in Fig. \ref{band} for $U$=2 eV, 
a value slightly
above $U_{cr}$ to make the gaps more clearly visible.  
For the ZZ case, the minimum
gap occurs at a corner of the zone that is not shown.  The dispersions going
away from the zone center of the uppermost two valence bands are 
entirely different for the two patterns,
being positive for ST but negative for ZZ.  The uppermost bands have primarily
Co2 $e_g^{\prime} (t_{2g})$, and not $a_g$, character.  
In spite of these differences in
dispersion through the zone, the orbital-projected density
of states, not shown, is extremely similar for the ZZ and ST bands
shown in Fig. \ref{band}.

We focus now on the observed ST pattern.  The gap occurs at the zone
boundary point Y along $k_y$. 
Although Co $3d$ states border the gap on both sides, this is an unusual $d-d$
charge transfer gap (not the usual $p-d$ case), with unoccupied Co1 minority
$a_g$ states above, and primarily Co2 $e_g^{\prime}$ states below the gap. 
The stronger dispersion of the upper Hubbard band above
the gap along  Y-$\Gamma$ compared to X-$\Gamma$ can be understood as follows.  
Electrons excited into the upper Hubbard reside in
the minority $a_g$ states on Co1, for example, a spin $\uparrow$ electron will
hop between Co1 ions with moments oriented $\downarrow$.  Propagating
in the $\hat y$ direction, it can hop through a single Co2 ion; in the $\hat x$
direction however, it must avoid the $U$ cost of hopping onto an oppositely
aligned Co1 ion, thus requiring hops through {\it two} Co2 ions before returning
to another Co1 $\downarrow$ ion, and its dispersion is reduced accordingly. 
Valence band holes introduced into the system
will occupy nonmagnetic Co2 $e_g^{\prime}$ states, while electrons will occupy
minority $a_g$ states on Co1.  

Since the crystal field ($t_{2g}\rightarrow e_g$) gap is $\sim$2 eV, the 
optical transitions in the IR for the magnetically disordered metallic phase
(T$>$T$_{c1}$) reflect $e_g^{\prime}\rightarrow a_g$ excitations, {\it i.e.}
transitions within the $t_{2g}$ complex.  Below the
metal-insulator transition at T$_{c2}$, the excitations across the gap
are to the upper
Hubbard band, and the main weight of these transitions -- the new 
(Co1 $\leftrightarrow$ Co2) 
charge-transfer type mentioned above -- is shifted up in
energy by only a few tens of meV.\cite{wang,hwang,lupi}  
This small shift is consistent with
the small bandwidth that we find for the upper Hubbard band. 

An interesting analogy with the AFM cuprate layer arises.
In this observed ST pattern Fig. \ref{cell}(b), the antiferromagnetic
arrangement has the bipartite, and 90$^{\circ}$, topology of spins characteristic
of the cuprate plane,
except with anisotropy of (Co$^{4+}$-Co$^{4+}$)
parallel and perpendicular hopping amplitudes $t_x, t_y$
and exchange couplings $J_x, J_y$.  Note that exchange $J_x$ is 
between near neighbors
while $J_y$ is between second Co neighbors.  There are however strong distinctions
to be made with cuprates.  In cuprates $U$ is 3-4 times larger, the effective
metal-metal near neighbor hopping is 2-3 times greater, and the 2 eV gap
typifies a robust Mott insulator.  In this cobaltate system the
tiny gap reflects a marginally insulating correlated state, and the low energy
excitations require three $d$ bands, versus the dominance of the 
single $d_{x^2-y^2}$
state in the cuprates.
It also seems that magnetic coupling cannot be treated in the usual Heisenberg
form, because the superexchange mechanism is not dominant and the
rather flimsy moments depend strongly on the type of magnetic
order.  (A spin Hamiltonian might be reasonable for treatment of spin waves
within a given ordered state.)  

Our results suggest a specific picture of the temperature evolution at $x$=0.5.
Noting that the FM ordered layers for $x\geq$0.75 are consistent with itinerant 
character and the $0.5 < x < 0.75$ regime with fluctuation-suppressed
magnetism, the magnetic ordering below T$_{c1}$ may be more of a spin density
wave (SDW) character which gaps some but not all of the Fermi surface; recall that for
the ST pattern the FM-AFM energy difference is very small at small U, and 
that even at U=0 there is a substantial difference in moments on Co1 and Co2,
that is, an SDW .  Several band structure studies have pointed
out nesting features in the paramagnetic Fermi 
surface.\cite{singh1,PRB04,michelle}  The challenge that this picture must
face is that the primary magnetic order is unchanged at the insulating
transition T$_{c2}$:  
Ga\v{s}parovic {\it et al.} find that the principal 
ordered moment grows with decreasing
temperature\cite{mit} continuously through the insulating transition at T$_{c2}$.
The additional order that results in a kink in $\chi_c$ at T$_{c2}$ has not
yet been elucidated, but our results are consistent with the prevailing
picture that disproportionation arises finally at T$_{c2}$.  The redistribution of spectral weight below T$_{c2}$
observed in optical experiments show differences (weight shifted to 20-30 
meV\cite{lupi} or 70-100 meV\cite{wang}), but they seem consistent with 
the correlated band structure of Fig. \ref{band} and particularly the narrowness
of the unoccupied band.

It can reasonably be asked whether the ground state of
this system should be considered as a
correlated insulator, as outlined above, or instead as perhaps a SDW
(at T$_{c1}$) -- CDW (at T$_{c2}$) system.  Balicas {\it et al.} have found
that, when an applied in-plane field increases beyond 25 T the conductivity
increases by a factor of two (a sort of insulator-metal transition), and
observation of magnetoresistance oscillations suggest the restoration of
part of the Fermi surface.\cite{balicas}  
Certainly the insulating phase is delicate.
However, the observation of a substantial ordered moment on Co1 
(0.25-0.34 $\mu_B$)\cite{mit,yokoi} and little or none on Co2 speaks for a charge
disproportion picture (into identifiable Co$^{4+}$ and Co$^{3+}$ moments)
and hence a correlated insulator below T$_{c2}$.  The observed value of the
Co1 moment is reduced somewhat from our calculated value of $0.5 \mu_B$,
as would be expected from two-dimensional fluctuations of a spin-half
moment already reduced substantially by hybridization\cite{mar} with O $2p$ states.   



We acknowledge helpful communications with C. Bernhard,
M. D. Johannes, J. Kune\v{s}, 
and D. J. Singh.
This work was supported by DOE grant DE-FG03-01ER45876 and DOE's
Computational Materials Science Network.  W.E.P. acknowledges the
support of DOE's SSAAP Program.


\begin{thebibliography}{10}
\bibitem{explain}We suspect AFM solutions for the IP-ST pattern should
  exist, but since this pattern is not observed we have not pursued
  this question.
\end{thebibliography}
\end{document}